\documentclass[%
 reprint,
superscriptaddress,
 amsmath,amssymb,
 aps,
]{revtex4-2}

\usepackage{graphicx}% Include figure files
\usepackage{dcolumn}% Align table columns on decimal point
\usepackage{bm}% bold math
\begin{document}

\preprint{APS/123-QED}

\title{Quantum walk on a ladder}

\author{Hira Ali}
\affiliation{%
 School of Natural Sciences, National University of Sciences and Technology,
Islamabad, Pakistan
}%
\author{M. Naeem Shahid}%
\affiliation{%
  School of Natural Sciences, National University of Sciences and Technology,
Islamabad, Pakistan
}%

\affiliation{%
National Centre for Physics (NCP), Shahdra Valley Road, Islamabad
44000, Pakistan
}%

\date{\today}% It is always \today, today,

\begin{abstract}
We study quantum walk on a ladder with combination of conventional
and split-step protocols. The two components of the walk resulting
from periodic boundary conditions can be made to have three kinds
of probability distributions. Two of these are the one-sided and alternative
ones. Using the differences between coin states of the individual
component, we can simulate the third case in which both components
have identical probability profiles. We also observe that the mutual
information transfer between two components of the walk is minimized
when this difference across the two sides of the ladder is maximized.
\end{abstract}

\maketitle

%\tableofcontents

\section{\label{sec:level1} Introduction}
The term quantum walk (QW) was first introduced in the seminal work
by Aharonov et al \citep{Aharonov_intro}. Since then, QWs have been
studied from computation point of view such as in decision trees \citep{Farhi_path,Farhi_decision},
cellular automata \citep{Meyer_CA}, quantum search algorithms \citep{Shenvi_SA,Ambainis_SA,Kendon_SA}
and universal quantum computing \citep{Childs_QC,Lovett_QC}. In addition,
QWs have been used as a tool to explore topological phases \citep{Karski_cold,Zahringer_ion,Kitagawa_split,Berry_topo,Wang_exp_walk,Feder_topo},
photosynthetic energy transfer \citep{Mohseni_photo}, Anderson localization
and decoherence \citep{Schreiber_decoh,Crespi_anderson,Edge_decoh}.
Several experiments with ions, atoms and photons \citep{Schmitz_ion,Broome_photon,Schreiber_photon,Alberti_exp_tom}
have also been performed to realize QWs and to simulate its impressions
on these phenomena \citep{Obuse_topo,Rakovszky_split}.

In continuous time quantum walk (CTQW), a quantum state is evolved
by unitary operator $U\left(t\right)={\rm exp}\left(-iHt\right)$
for some time $T$ \citep{Childs_qw_cw}, with the Hamiltonian $H$
governing the walk. This idea is inspired from classical Markov chains
where the role of $H$ is played by the adjacency matrix of the underlying
graph. Whereas, in discrete time quantum walk (DTQW) \citep{Meyer_CA},
a quantum coin is used to guide the walker along a graph. The size
of the wave packet after $n$ steps is given by \citep{Ahlbrecht_packet_width},
\begin{equation}
\left\langle \Delta x^{2}\right\rangle =n^{2}\left(1-\left|sin\left(\gamma/2\right)\right|\right)
\end{equation}
with $\gamma$ as the coin angle. This suggests the ballistic nature
of QW compared to classical one which is diffusive in nature as $\left\langle \Delta x^{2}\right\rangle =n$.
Moreover, it was also shown that DTQWs are faster than CTQWs \citep{coin_faster},
for further details please see Ref. \citep{Kempe_intro}, an excellent
review on the subject.

There is much work done on DTQWs on graphs \citep{Aharonov_graphs}
and 2D lattices \citep{Bru_cylinder}. In a single dimension, the
probability profile for quantum walk spreads at each step along the
line. In higher dimensions this distribution is further extended along
all direction depending on the protocol used for the walk. In a recent
article \citep{Bru_cylinder}, authors did analysis for a quantum
walker on a cylinder. They showed that the boundary conditions on the
closed side of cylinder implies several one dimensional walks characterized
by their coin angles. They used conventional walk protocol and marginal
probability as a measure. This protocol when used for the case of
ladder, keeps the walk on one side, see
Sec. IIA. On the other hand, in \citep{Omar_ent} authors considered
two walkers on a line where due to entanglement and relative phases
between the states influence the walk. Here, we explore DTQW on a ladder
where due to the boundary conditions the walk is split into effectively
two one dimensional components. The simple geometry of ladder lets
us observe and control the behavior of individual component as well
as the overall walk. Using split-step protocol,
we can give richer possibilities for the walk. We also used magnetization
like parameters discussed in Ref. \citep{Souza_mag} and thermodynamic-like quantities associated with the walk as in Ref. \citep{Romanelli_therm,ROMANELLI_therm1,Vallejo_therm}.

This article is organized as follows. In Sec. II, we describe the
setup for our model. In particular, we will show that our choice of the split-step protocol provides more freedom for the walker. In
Sec. III, we present our observation on the basis of few simple parameter
in analogy with paramagnetism. We finally conclude and summarize in
the last section.

\section{The choice of protocol}
The state of a quantum walker is typically labeled by its position
$\left|m\right\rangle $ and spin $\left|s\right\rangle $, degrees
of freedom. Initially this state $\left|s,m\right\rangle =\left|s\right\rangle \otimes\left|m\right\rangle $
can be set to any position (usually taken to be localized) and spin
(up, down or any mixture). Of course, these choices have effects on
the final probability distribution. The state is then evolved through
a given protocol for $N$ steps and the final state is projected over
all positions to get the probability distribution for the walker \citep{Kempe_intro}.

The conventional protocol for 1DQW is the flip of coin followed by
the shift operator as $\hat{U}^{{\rm con}}=\hat{T}\hat{C}\left(\gamma/2\right)$.
Here $\hat{C}$ is the coin operator with some angle $\gamma\in\left[0,\pi\right]$.
The general rotation of the coin allows for richer scenarios by changing
the mixing between up and down components and hence more control over
the walker. In this article we will use the coin introduced in the
experiments as in Refs. \citep{Karski_cold,Zahringer_ion} and for exploring topological
states \citep{Kitagawa_split},

\begin{equation}
\hat{C}\left(\gamma/2\right)=c\left|\uparrow\right\rangle \left\langle \uparrow\right|-s\left|\uparrow\right\rangle \left\langle \downarrow\right|+s\left|\downarrow\right\rangle \left\langle \uparrow\right|+c\left|\downarrow\right\rangle \left\langle \downarrow\right|,
\end{equation}

where $c=cos\left(\gamma/2\right),s=sin\left(\gamma/2\right)$. The
translation operator $\hat{T}$ decides the step direction by measuring
the spin of the walker,
\begin{equation}
\hat{T}=\sum_{m=-\infty}^{\infty}\left(\left|\uparrow\right\rangle \left\langle \uparrow\right|\otimes\left|m+1\right\rangle \left\langle m\right|+\left|\downarrow\right\rangle \left\langle \downarrow\right|\otimes\left|m-1\right\rangle \left\langle m\right|\right).
\end{equation}

Another choice for the walker on a line is to split the walk in two
half-steps with use of two coins. The unitary evolution for split-step
walk is $\hat{U}^{{\rm split}}=\hat{T}^{\downarrow}\hat{C}\left(\beta\right)\hat{T}^{\uparrow}\hat{C}\left(\alpha\right)$
where one component of the spin is moved and other is held after the
flip of first coin. Then the next coin is flipped and the process
is repeated for the other component. The half-shift operators are,

\begin{eqnarray}
T^{\uparrow} & = & \sum_{m=-\infty}^{\infty}\left(\left|\uparrow\right\rangle \left\langle \uparrow\right|\otimes\left|m+1\right\rangle \left\langle m\right|+\left|\downarrow\right\rangle \left\langle \downarrow\right|\otimes\left|m\right\rangle \left\langle m\right|\right),\\
T^{\downarrow} & = & \sum_{m=-\infty}^{\infty}\left(\left|\uparrow\right\rangle \left\langle \uparrow\right|\otimes\left|m\right\rangle \left\langle m\right|+\left|\downarrow\right\rangle \left\langle \downarrow\right|\otimes\left|m-1\right\rangle \left\langle m\right|\right).
\end{eqnarray}

The one dimensional protocol can be extended to two dimensional quantum
walks (2DQWs) with the coin operators replaced by the Groover coin
or the product of Hadamard coins.

This article uses one of the result given in the work by Bru et al
\citep{Bru_cylinder}, where they discussed the 2DQW on a cylinder
and showed that the boundary conditions implies a collection of 1DQWs
with various group velocities. In Fig. 1, it is shown that the use
of conventional protocol for both directions keeps the walker on one
side of the ladder as in alternative quantum walk (AQW).

\begin{figure}
\includegraphics[scale=0.3]{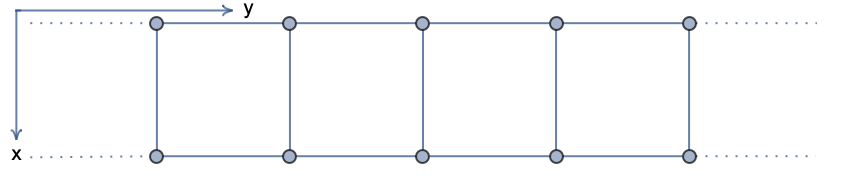}

\includegraphics[scale=0.4]{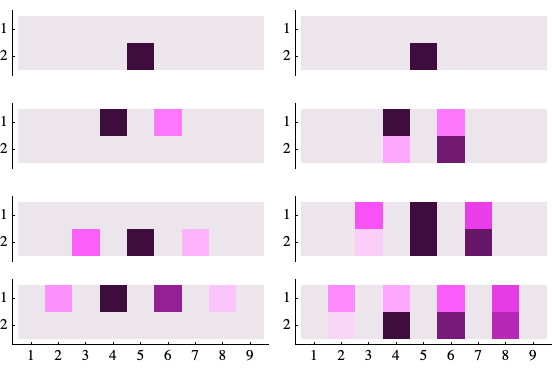}

\caption{}

First three step of the conventional (left) and split-step walk (right)
on a ladder. For conventional walk, the longer side coin is set to
$-\pi/4$ while for the shorter side it is $-\pi/8$. For split-step
$\alpha=-\pi/4$ and $\beta=-\pi/2$.
\end{figure}

Our aim is to keep the walker on both sides of the ladder. In figure
1, we show a simple ladder graph that is placed with its longer side
along $y$ axis and shorter side along $x$ axis. For the longer side
of the ladder, we choose the conventional protocol as $\hat{U}_{y}=\hat{T}\hat{C}\left(-\pi/4\right)$
with equal mixture of up and down spin components. However, if we
use the split walk then the walker can be kept on both sides of the
ladder. One step unitary evolution on the ladder can then be,

\begin{equation}
\hat{U}=\hat{U}_{y}^{{\rm split}}\hat{U}_{x}^{{\rm con}}=\hat{T}_{y}\hat{C}\left(\pi/4\right)\hat{T}_{x}^{\downarrow}\hat{C}\left(\beta/2\right)\hat{T}_{x}^{\uparrow}\hat{C}\left(\alpha/2\right).
\end{equation}

We now impose the boundary conditions on the shorter side of the ladder.
This leads to the quasi-momentum quantization along $x$ direction
as $k_{x}=n\pi$. The translation operator takes two distinct values
$+1$ or $-1$ as $e^{\pm ik_{x}}=\left(-1\right)^{n}$. This represents
two one dimensional walkers moving along $y$ direction. The unitary
operators for the two walkers are now,
\begin{eqnarray}
\hat{U_{1}} & = & \hat{T}\left(k\right)\hat{C}\left(\gamma_{1}/2\right),\\
\hat{U_{2}} & = & \hat{T}\left(k\right)\hat{C}\left(\gamma_{2}/2\right),
\end{eqnarray}

where $\gamma_{1}=\alpha+\beta-\pi/2$ and $\gamma_{2}=\gamma_{1}+\phi=\alpha-\beta+3\pi/2$.
Note that the phase difference between the coins is $\phi=\pm\left(2\beta-2\pi\right)$.

\section{Walk on the ladder}

A variety of probability distributions can be produced by setting
the coin angle to different values \citep{Panahiyan_stepcoin}. The
dependance of walk on coin angle has analogy with paramagnet in a
magnetic field \citep{Souza_mag}, where the role of the magnetic field
is replaced by the Bloch vector. As the angle of the coin is varied,
the probability of up and down spin components is changed, as shown
in Fig. 2. It can be seen from Eqs. (\ref{eq:dens_mat}) that the
difference of these probabilities is $1-\left|sin\left(\gamma/2\right)\right|$
for one dimensional walk. This parameter gives hint about the walker's
distribution in position space \citep{Souza_mag}.

In Fig. 2, the initial state is set at the center of the ladder on
one side while the spin is set in up state. The angle of coin is then
chosen as $0,\pi/4,\pi/2$ and $\pi$. At $\gamma=0$ and $\pi$,
this parameter takes its maximum and minimum values respectively and
the walk is classical-like. In former case, the walker moves without
any spread while for the latter it just oscillates around its initial
position. For the Hadamard, case the situation is like a fair classical
coin. The probabilities for both up and down components are equal
and hence the distribution is well spread over all points.

\begin{figure}
\includegraphics[scale=0.4]{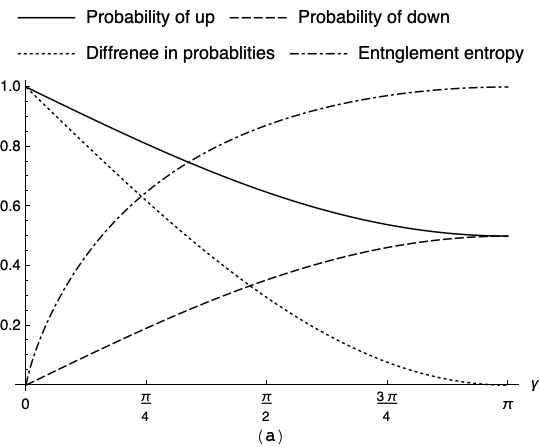}

\includegraphics[scale=0.5]{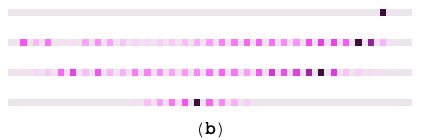}

\caption{}
a) Entropy and probabilities for a single walker with initial coin
state $\left(1,0\right)^{T}$. b) One dimensional walk for 31 steps
for coin angles $0,\pi/4,\pi/2$ and $\pi$ from top to bottom respectively
(darker colors indicate higher probabilities).
\end{figure}

We will now explore various walks on a ladder on the basis of these
parameters. Since the walk is decomposed effectively into two one
dimensional components, we can associate three such parameters. The
first two are the usual ones for each side and can be written down
immediately from Eqs. (\ref{eq:dens_mat}) as,

\begin{eqnarray}
M_{1} & = & 1-\left|sin\left(\gamma_{1}/2\right)\right|,\label{eq:m1}\\
M_{2} & = & 1-\left|sin\left(\gamma_{2}/2\right)\right|.\label{eq:m2}
\end{eqnarray}

For the probability differences across the ladder, we define a third
quantity as,

\begin{eqnarray}
M & = & 1-\frac{1}{2}\left|sin\left(\frac{\gamma_{1}}{2}\right)\right|-\frac{1}{2}\left|sin\frac{1}{2}\left(\frac{\gamma_{2}}{2}\right)\right|.\nonumber \\
 & = & \frac{1}{2}\left(M_{1}+M_{2}\right)
\end{eqnarray}

This turns out to be merely the average of $M_{1}$ and $M_{2}$.
Of course, this is also the difference between the diagonal components
of full density matrix.

\begin{figure}
\includegraphics[scale=0.4]{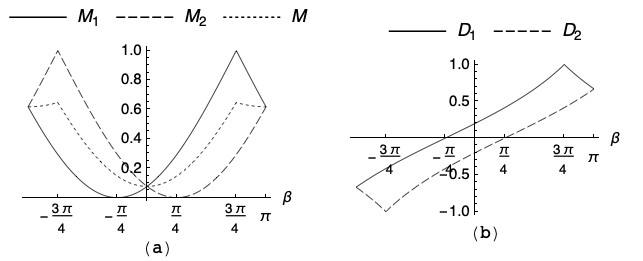}

\includegraphics[scale=0.5]{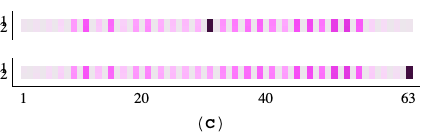}

\caption{}

a) $M_{1},M_{2}$ and $M$ vs $\beta$ for $\alpha=-\pi/4$. b) Discriminant
of the eigenvalues of density matrices for two components. c) Identical
walk for $\alpha=-\pi/4$ and $\beta=\pi/4,3\pi/4$.
\end{figure}

For any value of $\alpha$, when $\beta$ is an odd multiple of $\pm\pi$,
the walker is restricted to only one side of the ladder. It is also
evident from the periodicity of Eqs. (\ref{eq:m1}) and (\ref{eq:m2})
that all three parameters are equal at these points. However, at $\beta=0$,
the second coin is identity and the split-step walk is reduced to
alternative walk between two sides of the ladder, see Fig. 1. As $\beta$
is changed from $\pm\pi$, the walk starts to split on both sides of
the ladder until either $M_{1}$ or $M_{2}$ vanishes ($\alpha\pm\beta=-\pi/2$)
or gets maximized ($\alpha\pm\beta=\pi/2$). At these points, the
distribution on both sides of the ladder is identical as shown in
Fig 3. The whole walk is dominated by one component as the second
one is classical-like. These two cases are similar to 1DQW with coin
angles $0$ and $\pi$, as in Fig. 2, except that the walkers are
spread over all points. For the Hadamard cases when $\alpha=\pm\pi/2$,
such cases can not be observed as the difference between $M_{1}$
and $M_{2}$ is zero for all values of $\beta$ and none of the components
takes over. A summary of these results in given in Tab. 1.

\begin{figure}
\includegraphics[scale=0.25]{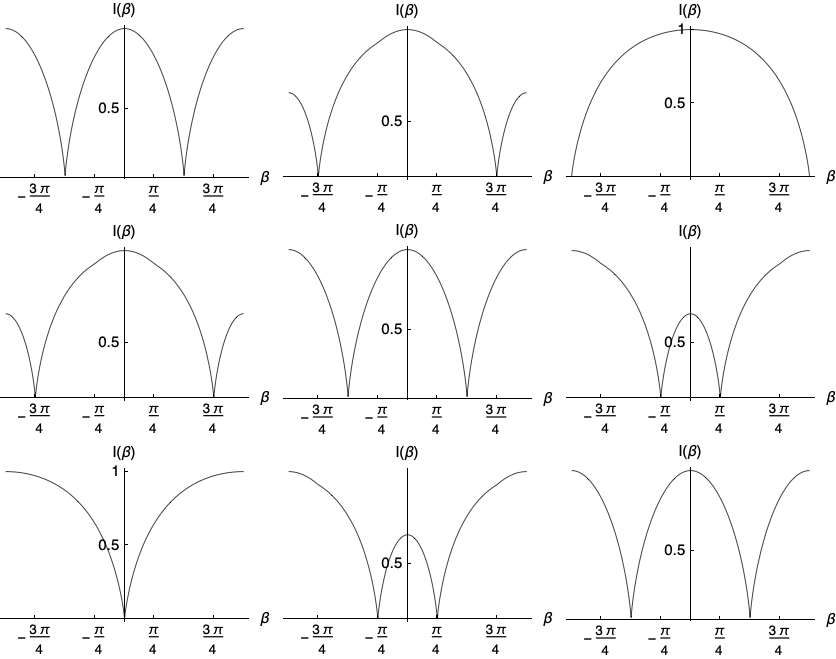}

\caption{Mutual information $I$ vs $\beta$ for $\alpha\in\left[-\pi,\pi\right]$
in $\pi/4$ intervals.}

\end{figure}

\begin{table}[b]
\caption{Summary of various walk pattern.}
\begin{ruledtabular}
\begin{tabular}{lll}
$\alpha\pm\beta$ & Magnetization & Walk patterns\\
\colrule
$\alpha\pm0$ & $M_{1}=M_{2}=M$ & Alternating
\\
$\alpha\pm\pi$ & $M_{1}=M_{2}=M$ & One-sided
\\ 
$-\pi/2$ & $M_{1}$or $M_{2}=0$ & Identical
\\
$\pi/2$ & $M_{1}$or $M_{2}$ is maximized & Identical
\\ 
\end{tabular}
\end{ruledtabular}

\end{table}

We would like to remark that these arguments can be based on other
quantities. For example, the discriminants in the eigenvalues of density
matrices are of the form,

\begin{equation}
D=\frac{\left|cos\left(\gamma/4\right)\right|-\left|sin\left(\gamma/4\right)\right|}{\left|cos\left(\gamma/4\right)\right|+\left|sin\left(\gamma/4\right)\right|}.
\end{equation}

In Fig 3. we have plotted $D_{1}$ and $D_{2}$ for both components
where one can note the above mention cases. Indeed, various thermodynamic-like
properties can be assigned to a quantum walker, which are related
to these variables \citep{Romanelli_therm,ROMANELLI_therm1,Vallejo_therm}.

We finally use mutual quantum information \citep{Xue_ent} for the
two components,
\begin{equation}
I=S\left(\rho_{1}\right)+S\left(\rho_{2}\right)-S\left(\rho\right),
\end{equation}

to distinguish these two types identical distributions. Here $S\left(\rho_{1}\right)$,
$S\left(\rho_{2}\right)$ and $S\left(\rho\right)$ are the von Neumann
entropies associated with each component and overall walk, respectively.
The zeros of this expression indicate the angles at which the two
components are independent. As an example, $\alpha=-\pi/4$ and $\beta=\pm3\pi/4$
corresponds to two independent walks while at $\beta=0$, both distributions
are fully dependent on each other, see Fig 3 and Fig. 4. This may
be the case as the walk is alternative at this value and hence each
component is fully determined by the other.

\section{Conclusion}

We considered the simple situation for a quantum walker moving on
a ladder. Using a specific mixed protocol for the walk, in contrast
to conventional protocols, gives an extra option for the walker to be
on both siders of the ladder. The analysis is done on the basis of
probability differences of up and down spin components. We have also
shown that for particular choices of two angles, the components can
have almost similar probability profiles. As the walk evolves from
one side of the ladder to the other, the mutual information is transferred
from one component to the other. We have shown that this transfer
is zero when the probability difference between up and down spin components
across the ladder is maximum. These cases are simpler for a ladder
but generalizing to a cylinder will need more angles and hence may
not be feasible.

\begin{acknowledgments}
We would like to thank Aeysha Khalique for useful discussions at the
initial stage of this work.

\end{acknowledgments}

\appendix

\section{Density matrices for the walk}

The density matrix encodes the information about final state probabilities
and hence the entropies for the quantum walk \citep{Endrejat_ent,Venegas_ent,Omar_ent,Pathak_ent,Liu_ent,Annabestani_ent,Xue_ent}.
The evolutions of the two walkers can then be related to their corresponding
density matrices. Since both walks are effectively one dimensional
and differs only by coin angle, we will compute the density matrix
for one walker and replace the coin angle for the other. For simplicity
we choose the initial state of the walkers to be $\left|\psi\right\rangle _{0}=\left(1,0\right)^{T}$which
corresponds to $\theta=0$ and $\phi=0$ on the Bloch sphere. The
results can be easily generalized for a general state \citep{Nayak_denstiy,Abal_density}.

The state $\left|\psi\right\rangle _{0}$ is then evolved $n$ times
by the corresponding unitary operator which can be written using
spectral decomposition as,

\begin{equation}
U^{n}=e^{-i\omega n}\left|e^{+}\right\rangle \left\langle e^{+}\right|+e^{i\omega n}\left|e^{-}\right\rangle \left\langle e^{-}\right|,
\end{equation}

where $cos\omega_{\pm}=cos\left(\gamma/2\right)cosk$ is related to
the eigenvalues $\lambda^{\pm}=e^{\pm i\omega}$of the unitary operator.
The eigenvectors $\left|e^{\pm}\right\rangle $ of the unitary operator
are given by,
\begin{equation}
\left|e^{\pm}\right\rangle =\frac{1}{N_{\pm}}\left(\begin{array}{c}
u\\
-v^{\pm}+w
\end{array}\right),
\end{equation}

with, 
\begin{eqnarray}
u & = & e^{ik},\nonumber \\
v & = & e^{ik}cot\left(\gamma/2\right),\nonumber \\
w & = & cosec\left(\gamma/2\right)e^{\pm i\omega},
\end{eqnarray}

and normalization,
\begin{equation}
N^{\pm}=\frac{sin\left(\gamma/2\right)}{\sqrt{2\left[sin^{2}\omega\mp cos\left(\gamma/2\right)sinksin\omega\right]}}.
\end{equation}

The evolved state $\left|\psi\right\rangle =\left(a\left(k\right),b\left(k\right)\right)^{T}$
is then used to compute the density matrix. In the large $n$ limit,
it has the following elements,
\begin{eqnarray}
\rho^{11} & = & 1-\frac{1}{2}\left|sin\left(\gamma/2\right)\right|,\nonumber \\
\rho^{22} & = & 1-\rho^{11},\nonumber \\
\rho^{12} & = & \rho^{21}=-\frac{1}{2}\left[1-sin\left(\gamma/2\right)\right]tan\left(\gamma/2\right).\label{eq:dens_mat}
\end{eqnarray}

Here we used the fact that ${\rm tr}\left(\rho\right)=1$, as a consequence
of probability conservation.

With definition, $\Delta=\text{\ensuremath{\rho^{11}\rho^{22}}}-\left|\rho^{12}\right|^{2}$
the eigenvalues of the density matrix can be computed as,

\begin{eqnarray}
\Lambda^{\pm} & = & \frac{1}{2}\left(1\pm\sqrt{1-4\Delta}\right)\nonumber \\
 & = & \frac{1}{1+tan\left(\gamma/2\right)},\frac{1}{1+cot\left(\gamma/2\right)}
\end{eqnarray}

For each component we can now replace $\gamma$ by $\gamma_{1}$ and
$\gamma_{2}$. The full density matrix is then the average of these
density matrices. Finally the entanglement entropies can be computed
using,

\begin{equation}
S=-\Lambda^{-}{\rm log}_{2}\Lambda^{-}-\Lambda^{+}{\rm log}_{2}\Lambda^{+},
\end{equation}

for each component and the whole walk.

\bibliography{walk}% Produces the bibliography via BibTeX.

\end{document}